\documentclass[journal,onecolumn]{IEEEtran}
\usepackage[ruled,linesnumbered]{algorithm2e}

\renewcommand{\IEEEQED}{\IEEEQEDclosed}
\usepackage{color}

\usepackage{subfigure}
\usepackage{graphicx,cite,epsfig,amssymb,amsmath,multirow,lettrine,flushend,extarrows}

\begin{document}



\title{Deep Reinforcement Learning for Smart Home Energy Management}

\author{{Liang~Yu,~\IEEEmembership{Member,~IEEE}, Weiwei Xie, Di Xie, Yulong Zou,~\IEEEmembership{Senior Member,~IEEE}, Dengyin Zhang,\\ Zhixin Sun, Linghua Zhang, Yue~Zhang,~\IEEEmembership{Senior Member,~IEEE}, Tao~Jiang,~\IEEEmembership{Fellow,~IEEE}}
\thanks{
\newline L. Yu, W. Xie, D. Xie, Y. Zou, Z. Sun, L. Zhang are with Key Laboratory of Broadband Wireless Communication and Sensor Network Technology of Ministry of Education, Nanjing University of Posts and Telecommunications, Nanjing 210003, P. R. China.
\newline D. Zhang is with Jiangsu Key Laboratory of Broadband Wireless Communication and Internet of Things, School of Internet of Things, Nanjing University of Posts and Telecommunications, Nanjing 210003, P. R. China. \newline
Y. Zhang is with the Department of Engineering, University of Leicester, Leicester LE1 7RH, U.K.\newline
T. Jiang is with Wuhan National Laboratory for Optoelectronics, School of Electronic Information and Communications, Huazhong University of Science and Technology, Wuhan 430074, P. R. China.  \newline
}}

\markboth{IEEE Internet of Things Journal,~Vol.~XX, No.~XX, Month~2019}%
{Liang \MakeLowercase{\textit{et al.}}: Deep Reinforcement Learning for Smart Home Energy Management}

\maketitle

\begin{abstract}
In this paper, we investigate an energy cost minimization problem for a smart home in the absence of a building thermal dynamics model with the consideration of a comfortable temperature range. Due to the existence of model uncertainty, parameter uncertainty (e.g., renewable generation output, non-shiftable power demand, outdoor temperature, and electricity price) and temporally-coupled operational constraints, it is very challenging to design an optimal energy management algorithm for scheduling Heating, Ventilation, and Air Conditioning (HVAC) systems and energy storage systems in the smart home. To address the challenge, we first formulate the above problem as a Markov decision process, and then propose an energy management algorithm based on Deep Deterministic Policy Gradients (DDPG). It is worth mentioning that the proposed algorithm does not require the prior knowledge of uncertain parameters and building thermal dynamics model. Simulation results based on real-world traces demonstrate the effectiveness and robustness of the proposed algorithm.
\end{abstract}


\section{Introduction}\label{s1}
As a next-generation power system, smart grid is typified by an increased use of information and communications technology (e.g., Internet of Things) in the generation, transmission, distribution, and consumption of electrical energy. In smart grid environment, there are many opportunities for saving the energy cost of smart homes, which are evolved from traditional homes by adopting three components, i.e., the internal networks, intelligent controls, and home automations\cite{Wu2017}. For example, dynamic electricity prices could be utilized to reduce energy cost by scheduling Energy Storage Systems (ESS) and thermostatically controllable loads intelligently. As one kind of thermostatically controllable loads, Heating, Ventilation, and Air Conditioning (HVAC) systems consume about 40\% of total energy in a household\cite{Afram2016}, which results in energy cost concerns for smart home owners. Since the primary purpose of HVAC systems is to maintain thermal comfort for the occupants, it is of great importance to optimize the energy cost of smart homes without sacrificing thermal comfort.

In this paper, we investigate an energy optimization problem for a smart home with renewable energies, ESS, HVAC systems, and non-shiftable loads (e.g., televisions) in the absence of a building thermal dynamics model. To be specific, our objective is to minimize the energy cost of the smart home during a time horizon with the consideration of a comfortable indoor temperature range. However, it is very challenging to achieve the above aim due to the following reasons. Firstly, it is often intractable to obtain accurate dynamics of indoor temperature, which can be affected by many factors\cite{Wei2017}. Secondly, it is difficult to know the statistical distributions of all combinations of random system parameters (e.g., renewable generation output, power demand of non-shiftable loads, outdoor temperature, and electricity price). Thirdly, there are temporally-coupled operational constraints associated with ESS and HVAC systems, which means that the current action would affect the future decisions. To address the above challenge, we propose a Deep Deterministic Policy Gradients (DDPG) based energy management algorithm, which can make decision about ESS charging/discharging power and HVAC input power simply based on the current observation information.

The main contributions of this paper are summarized as follows.
\begin{itemize}
  \item We investigate an energy cost minimization problem for smart homes in the absence of a building thermal dynamics model with the consideration of a comfortable temperature range, energy exchange between the smart home and the utility grid, ESS charging/discharging, HVAC input power adjustment, and parameter uncertainties. Then, we reformulate the problem as a Markov Decision Process (MDP), where environment state, action and reward function are designed.
  \item We propose an energy management algorithm to jointly schedule ESS and HVAC systems based on DDPG. Since the proposed algorithm makes decision simply based on the current environment state, it does not require prior knowledge of uncertain parameters and building thermal dynamics model.
  \item Extensive simulation results based on real-world traces show that the proposed algorithm can save energy cost by 8.10\%-15.21\% without sacrificing thermal comfort when compared with two baselines. Moreover, the robustness testing shows that the proposed algorithm has the potential of providing a more efficient and practical tradeoff between maintaining thermal comfort and reducing energy cost than an ``optimal" strategy.
\end{itemize}

The remainder of this paper is organized as follows. In Section \ref{s2}, we introduce related works. In Section~\ref{s3}, system model and problem formulation are given. Then, we propose a DDPG-based energy management algorithm in Section~\ref{s4} and its effectiveness is verified by simulation results in Section~\ref{s5}. Finally, we make a conclusion and discuss the future work in Section~\ref{s6}.

\section{Related Works}\label{s2}
There have been many studies on energy cost and/or thermal comfort in smart homes. Due to the space limitation, we mainly focus on joint energy cost and thermal comfort management in smart homes\cite{Angelis2013,Fan2016,Zhang2016,Pilloni2018,LiangYuTSG2019}. The approaches proposed in these studies can be generally classified into two categories, i.e., model-based approaches and model-free based approaches. To be specific, model-based approaches are designed based on the model information about thermal dynamics of the environment\cite{Shad2015}\cite{Kara2015}. By contrast, model-free based approaches are designed without requiring the above-mentioned information.

\subsection{Model-based approaches}
In \cite{Angelis2013}, Angelis \emph{et al.} presented a home energy management approach to minimize the energy cost related to task execution, energy storage, energy selling and heat pump without violating the given comfortable temperature range and other constraints. In \cite{Fan2016}, Fan \emph{et al.} proposed an online home energy management scheme to minimize the energy cost associated with electric water heaters and HVAC systems with the consideration of indoor temperature ranges. In \cite{Zhang2016}, Zhang \emph{et al.} developed a home energy management strategy to minimize energy cost related to the HVAC load and deferrable loads without violating the given comfortable temperature range. In \cite{Pilloni2018}, Pilloni \emph{et al.} proposed a Quality of Experience (QoE)-aware smart home energy management system to save energy cost while minimizing the annoyance perceived by users. In \cite{LiangYuTSG2019}, Yu \emph{et al.} proposed an online home energy management algorithm to minimize the sum of energy cost and thermal discomfort cost (Here, thermal discomfort cost is the function of temperature deviation between indoor temperature and the comfortable temperature level). In \cite{Franceschelli2018}, Franceschelli \emph{et al.} proposed a heuristic approach to optimize the peak-to-average power ratio of a large population of thermostatically controlled loads considering comfortable temperature ranges. Although some advances have been made in the above-mentioned works, their approaches need to model building thermal dynamics with simplified mathematical models, e.g., Equivalent Thermal Parameters (ETP) model.

\subsection{Model-free based approaches}
Since it is very challenging to develop a building thermal dynamics model that is both accurate and efficient enough for HVAC control, some recent works have considered to use real-time data for HVAC control\cite{LuTSG2019,Ruelens2017,Canteli2019}. For example, Lu \emph{et al.} in \cite{LuTSG2019} proposed an energy management scheme to minimize the sum of electricity cost and user dissatisfaction cost associated with wash machines and HVAC loads based on multi-agent reinforcement learning and artificial neural network approach. In \cite{Ruelens2017}, Ruelens \emph{et al.} proposed a residential demand response method to minimize energy cost with the consideration of temperature range based on batch reinforcement learning. Although reinforcement learning based methods in \cite{LuTSG2019,Ruelens2017,Canteli2019} do not require the prior knowledge of building thermal dynamics model, they are known to be unstable or even to diverge when a nonlinear function approximator (e.g., a neural network) is used to represent the action-value function\cite{Mnih2015}. To efficiently handle large and continuous state space, deep reinforcement learning (DRL) has been presented and shown successful in playing Atari and Go games\cite{Mnih2015}. In \cite{Wei2017}, Wei \emph{et al.} proposed a DRL-based method for building HVAC control, which can reduce energy cost while maintaining the desired indoor temperature range. In \cite{Gao2019}, Gao \emph{et al.} presented a DRL-based thermal comfort control method to minimize energy consumption and thermal discomfort. In \cite{Zhang2018}, Zhang \emph{et al.} conducted real-life implementation and evaluation of a DRL-based control method for a radiant heating system, which optimizes energy demand and thermal comfort. In \cite{Valladares2019}, Valladares \emph{et al.} proposed a DRL-based thermal comfort and indoor air control algorithm. In \cite{Wan2018}, Wan \emph{et al.} proposed a DRL-based algorithm to minimize the energy cost of a smart home with battery energy storage. Although some model-free methods have been proposed in above-mentioned studies, none of them can be applicable to the coordination between ESS and HVAC systems in smart homes. To deal with this problem, we develop a DDPG-based energy management algorithm in this paper.

\section{System Model And Problem Formulation}\label{s3}
\begin{figure}[!htb]
\centering
\includegraphics[scale=0.87]{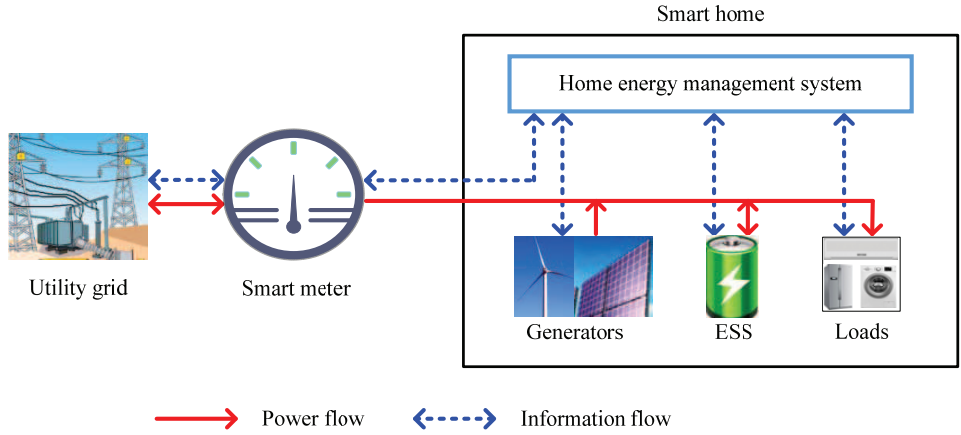}
\caption{Illustration of a smart home.}\label{fig_1}
\end{figure}

The smart home considered in this paper is shown in Fig.~\ref{fig_1}, where distributed generators, ESS, loads, and home energy management system (HEMS) could be identified. Distributed generators could be solar panels or wind generators. ESS could be lead-acid batteries or lithium-ion batteries, which can reduce net-energy demand from main grids by storing excess renewable energies locally and are very important for implementing nearly-zero energy buildings in the future\cite{Nousdilis2018}. At present, ESS costs are very high (e.g., around $450\$/\text{kWh}$), which means that installing ESS in a smart home is not very economical. However, ESS costs are dropping rapidly with the development of technology and are predicted to drop below $100\$/\text{kWh}$ within the next decade. As a result, the profitability of adopting ESS will gradually increase. Therefore, we consider ESS in the model of the smart home. Loads in a smart home can be generally divided into several types, e.g., non-shiftable loads, shiftable and non-interruptible loads, and controllable loads\cite{Yousefi2019}. To be specific, power demands of non-shiftable loads (e.g., televisions, microwaves, and computers) must be satisfied completely without delay. As for shiftable and non-interruptible loads (e.g., washing machines), their tasks can be scheduled to a proper time but can not be interrupted. In contrast, controllable loads (e.g., HVAC systems, heat pumps, and electric water heaters) can be controlled to flexibly adjust their operation times and energy usage quantities by following some operational requirements, e.g., temperature ranges. In this paper, we mainly focus on non-shiftable loads and thermostatically controlled loads\cite{Ruelens2017}. As for thermostatically controlled loads, HVAC systems are considered since they consume about 40\% of the total energy in a smart home\cite{Afram2016}. Suppose that the HEMS operates in slotted time, i.e., $t\in [1,T]$, where $T$ is the total number of time slots. For simplicity, the duration of a time slot $\Delta t$ is normalized to a unit time (e.g., one hour) so that power and energy could be used equivalently. In each time slot, the HEMS makes continuous decision on ESS charging/discharging power and HVAC input power according to a set of available information (e.g., renewable generation output, non-shiftable power demand, outdoor temperature, and electricity price), with the aim of minimizing the energy cost of the smart home while maintaining the comfortable temperature range in the absence of the building thermal dynamics model. In the following parts, models associated with ESS and HVAC systems are provided. Then, an energy cost minimization problem is formulated. Next, we reformulate it as a MDP due to the difficulty of solving the minimization problem.

\subsection{ESS Model}\label{s3_1}
Let $B_t$ be the stored energy in the ESS at time slot $t$. Then, the ESS storage dynamics model is given by
\begin{equation}\label{f_1}
B_{t + 1}=B_t + \eta_c c_t+\frac{d_t}{\eta_d},~\forall~t,
\end{equation}
where ${\eta_c} \in \left( {0,1} \right]$ and ${\eta_d} \in \left( {0,1} \right]$ are the charging and discharging efficiency coefficients, respectively; $c_t$ and $d_t$ are ESS charging power and discharging power, respectively. Here, $c_t$ and $d_t$ are assigned with different signs (i.e., $c_t\geq 0$ and $d_t\leq 0$), which contributes to the design of action in Section II-F.

Since ESS cannot be charged above its capacity $B^{\max}$ or discharged below the minimal energy level $B^{\min}$, we have
\begin{equation}\label{f_2}
B^{\min} \le {B_t} \le B^{\max},~\forall~t.
\end{equation}

Due to the existence of ESS charging and discharging rate limitations, we have
\begin{align}\label{f_3}
0\leq c_t\leq c^{\max},~\forall~t,\\
-d^{\max}\leq d_t\leq 0,~\forall~t,
\end{align}
where $c^{\max}$ and $d^{\max}$ are maximum charging and discharging power of the ESS, respectively.

To avoid the simultaneous ESS charging and discharging, we have
\begin{align}\label{f_4}
c_t\cdot d_t=0,~\forall~t.
\end{align}

\subsection{HVAC Model}\label{s3_2}
The HVAC system can be dynamically adjusted to maintain thermal comfort of the occupants in the smart home. Since thermal comfort depends on many factors (e.g., air temperature, mean radiant temperature, relative humidity, air speed, clothing insulation, and metabolic rate), its representation is very complex. In existing studies, many modeling approaches and parameter measurement methods associated with thermal comfort have been developed\cite{Gao2019}\cite{Yang2019,Cheng2019,Cheng2017,Wang2018,Li2018,Yang2018,YuJIOT2017}. Similar to \cite{Wei2017,Angelis2013,Fan2016,Zhang2016}, this paper uses a comfortable temperature range as the representation of thermal comfort for simplicity, i.e.,
\begin{align} \label{f_5}
T^{\min}\leq T_t\leq T^{\max},~\forall~t,
\end{align}
where $T^{\min}$ and $T^{\max}$ are the minimum and maximum comfort level, respectively.

In this paper, we consider an HVAC system with inverter in the smart home, i.e., the HVAC system can adjust its input power $e_t$ continuously\cite{LiangYuTSG2019}. Suppose $e^{\max }$ be the rating power of the HVAC system, we have
\begin{equation}\label{f_6}
0 \le {e_t} \le {e^{\max}},~\forall~t.
\end{equation}

\subsection{Power Balancing}\label{s3_3}
To keep the power balance in the smart home, the aggregated power supply should be equal to the served power demand. Then, we have
\begin{equation}\label{f_7}
g_t+p_t-d_t=b_t+e_t+c_t,\forall~t,
\end{equation}
where $g_t$, $p_t$, $b_t$ are power drawn from the utility grid, renewable generation output, and non-shiftable power demand, respectively. If $g_t<0$, it means that energy form the smart home will be sold to the utility grid. Otherwise, the smart home will purchase energy from the utility grid.

\subsection{Cost Model}\label{s3_4}
Let $v_t$ and $u_t$ be the buying and selling price of energy, respectively. Then, the energy cost of the smart home at time slot $t$ can be calculated by
\begin{equation}\label{f_8}
C_{1,t} = {({{{v_t} - {u_t}} \over 2}\left| {{g_t}} \right| + {{{v_t} + {u_t}} \over 2}{g_t})},~\forall~t,
\end{equation}
where the intuition behind \eqref{f_8} is that just one variable $g_t$ is needed to reflect the behavior of electricity buying or selling. For example, when $g_t\geq 0$, $C_{1,t}=v_tg_{t}$. For the case $g_t<0$, $C_{1,t}=u_tg_{t}$.

It is well known that frequent discharging or charging would do harm to the lifetime of the ESS. To capture this phenomenon, ESS depreciation cost at time slot $t$ is introduced as follows\cite{ArxivXu2019}
\begin{equation}\label{f_9}
C_{2,t} =\psi(\left| c_t \right| + \left| d_t \right|),~\forall~t,
\end{equation}
where $\psi$ denotes ESS depreciation coefficient in $\$/\text{kW}$.

\subsection{Total Energy Cost Minimization Problem}\label{s3_5}
Based on the above-mentioned models, we can formulate a total energy cost minimization problem as follows,
\begin{subequations}\label{f_10}
\begin{align}
(\textbf{P1})~&\min~\sum\limits_{t=1}^{T} \mathbb{E}\{C_{1,t}+C_{2,t}\}  \\
s.t.&~(1)-(8),
\end{align}
\end{subequations}
where the expectation operator $\mathbb{E}$ is taken over the randomness of the system parameters (i.e., renewable generation output $p_t$, non-shiftable power demand $b_t$, outdoor temperature $T_t^{\text{out}}$, and buying/selling electricity prices $v_t/u_t$) and the possibly random control actions (i.e., the amount of energy exchange between the smart home and the utility grid $g_t$, ESS charging/discharging power $c_t/d_t$, and HVAC input power $e_t$) at each time slot.

It is very challenging to solve \textbf{P1} due to the following reasons. Firstly, it is often intractable to obtain accurate dynamics of indoor temperature $T_t$, which can be affected by many factors\cite{Wei2017}, e.g., building structure and materials, surrounding environment (e.g., ambient temperature, humidity, and solar radiation intensity), and internal heat gains from occupants, lighting systems and other equipments. Secondly, it is very difficult to know the statistical distributions of all combinations of random system parameters. Thirdly, there are temporally-coupled operational constraints associated with ESS and HVAC systems, which means that the current action would affect future decisions. To handle the ``time-coupling" property, typical methods are based on dynamic programming\cite{LiangYuTSG2019}, which suffers from ``the curse of dimensionality" problem. In this paper, we provide a way of solving \textbf{P1} without requiring the dynamics of indoor temperature and prior knowledge of random system parameters. In particular, we reformulate the above-mentioned sequential decision making problem as a MDP problem. Then, we develop a DDPG-based energy management algorithm for the problem.

\subsection{MDP Formulation}\label{s3_6}
In the smart home, the indoor temperature at next time slot is only determined by the indoor temperature, HVAC power input, and environment disturbances (e.g., outdoor temperature and solar irradiance intensity) in the current time slot\cite{Zhang2016}\cite{Pilloni2018}\cite{Constantopoulos1991}\cite{Thatte2012}. Moreover, the ESS energy level at next time slot just depends on the current energy level and current discharging/charging power according to \eqref{f_1}, which is independent of previous states and actions. Thus, both of ESS scheduling and HVAC control can be regarded as a MDP. In the following parts, we will formulate the sequential decision making problem associated with smart home energy management as a MDP. It is worth noting that the MDP formulation is an approximation description of the smart home energy management problem since some components of the environment state may be not Markovian in practice, e.g., renewable generation output and electricity price. According to existing works\cite{Mnih2015}\cite{Zhangchong2018}, even though the environment is not strictly MDP, the corresponding problem can still be solved by reinforcement learning based algorithms empirically, which is also validated by simulation results in this paper. For non-Markovian environment, many approaches could be adopted to improve the performance of reinforcement learning based algorithms, e.g., approximate state\cite{Zhangchong2018}\cite{Sutton2018}, recurrent neural networks\cite{Schmidhuber1990}, gated end-to-end memory policy networks\cite{Perez2017}, and eligibility traces\cite{Sutton2018}.

\begin{figure}[!htb]
\centering
\includegraphics[scale=0.783]{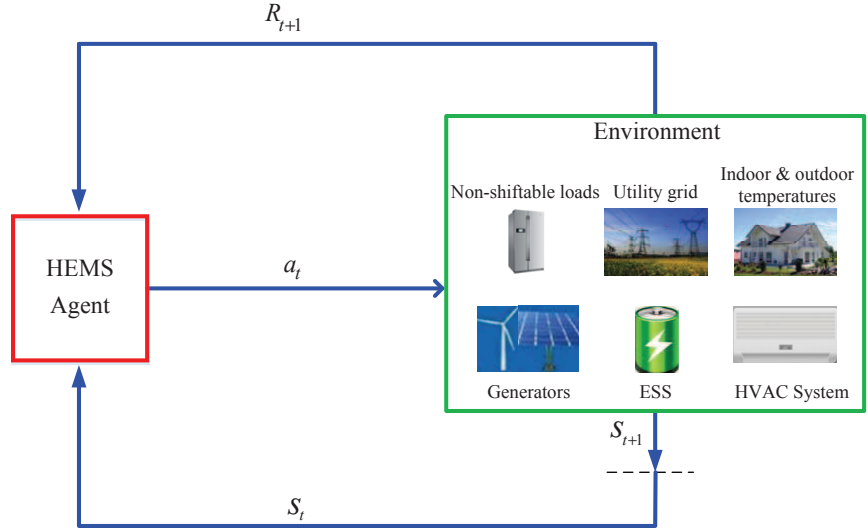}
\caption{The agent-environment interaction in the MDP.}\label{fig_2}
\end{figure}

A discounted MDP is formally defined as a five-tuple $M=(\mathcal{S},\mathcal{A},\mathcal{P},\mathcal{R},\gamma)$, where $\mathcal{S}$ is the set of environment states and $\mathcal{A}$ is the set of actions. $\mathcal{P}:~\mathcal{S}\times \mathcal{A}\times \mathcal{S}\rightarrow [0,1]$ is the transition probability function, which models the uncertainty in the evolution of states of the system based on the action taken by the agent\cite{Padakandla2019}. $\mathcal{R}:~\mathcal{S}\times \mathcal{A}\rightarrow \mathbb{R}$ is the reward function and $\gamma\in [0,1]$ is a discount factor. In this paper, the agent denotes the learner and decision maker (i.e., HEMS agent), while the environment comprising many objects outside the agent (e.g., renewable generators, non-shiftable loads, ESS, the HVAC system, utility grid, indoor/outdoor temperature). The interaction between the agent and the environment can be depicted by Fig.~\ref{fig_2}, where the HEMS agent observes environment state $\boldsymbol{s}_{t}$ and takes action $\boldsymbol{a}_t$. Then, environment state becomes $\boldsymbol{s}_{t+1}$ and the reward $R_{t+1}$ is returned. In the following parts, we will design key components of the MDP, including environment state, action and reward function.

\subsubsection{Environment State}
The environment state consists of seven kinds of information, i.e., renewable generation output $p_t$, non-shiftable power demand $b_t$, ESS energy level $B_t$, outdoor temperature $T_t^{\text{out}}$, indoor temperature $T_t$, buying electricity price $v_t$, and time slot index in a day $t'$ ($t'=\mod(t,24)$). Since selling electricity price $u_t$ is typically related to buying electricity price $v_t$ (e.g., $u_t=\delta v_t$\cite{liangyu2015,liangTSG2018,zhangyu2013}, $\delta$ is a constant), $u_t$ is not selected as a part of the environment state. For brevity, $\boldsymbol{s}_t$ is adopted to describe the environment state, i.e., $\boldsymbol{s}_t=(p_t,~b_t,~B_t,T_t^{\text{out}},T_t,v_t,t')$.

\subsubsection{Action}
The aim of HEMS agent is to optimally decide the amount of energy exchange between the smart home and the utility grid (i.e., $g_t$), ESS charging power (i.e., $c_t$), ESS discharging power (i.e., $d_t$), and HVAC input power $e_t$. After $c_t$, $d_t$, and $e_t$ are jointly decided, $g_t$ can be known immediately according to \eqref{f_7}. Therefore, the action of the MDP consists of ESS charging/discharging power $c_t/d_t$ and HVAC input power $e_t$. Since adopting $c_t$ and $d_t$ simultaneously would complicate the design of the energy management algorithm, we use just one variable $f_t$, where the range of $f_t$ is $[-d^{\max},~c^{\max}]$. When $f_t\geq 0$, $c_t=f_t$ and $d_t=0$. When $f_t\leq0$, $c_t=0$ and $d_t=f_t$. Therefore, the constraints (3)-(5) could be guaranteed. To guarantee the feasibility of (1)-(2), $0\leq c_t\leq \min\{c^{\max},\frac{B^{\max}-B_t}{\eta_c}\}$ when $f_t\geq 0$, and $\min\{-d^{\max},(B^{\min}-B_t)\eta_d\}\leq d_t \leq 0$ when $f_t\leq0$. According to \eqref{f_5}, the range of $e_t$ is $[0,~e^{\max}]$. When indoor temperature $T_t$ is lower than $T^{\min}$, $e_t$ should be zero for avoiding further temperature deviation. Similarly, when $T_t>T^{\max}$, the feasible $e_t$ should be nonnegative. For brevity, $\boldsymbol{a}_t$ is used to describe the action, i.e., $\boldsymbol{a}_t=(f_t,~e_t)$.

\subsubsection{Reward}
According to the MDP theory in \cite{Sutton2018}, the transition of the environment state from $\boldsymbol{s}_{t-1}$ to $\boldsymbol{s}_{t}$ could be triggered by the execution of $\boldsymbol{a}_{t-1}$. Finally, the reward $R_t$ will be obtained. Since the aim of the agent is to minimize the total energy cost while maintaining the comfortable temperature range, the corresponding reward consists of three parts, namely the penalty for the energy consumption of the HVAC system, the penalty for ESS depreciation, and the penalty for temperature deviation. Since the energy cost of the HVAC system at slot $t-1$ is $C_{1,t-1}$, the first part of $R_t$ can be represented by $-C_{1,t-1}(\boldsymbol{s}_{t-1},\boldsymbol{a}_{t-1})$. Similarly, the second part of $R_t$ can be described by $-C_{2,t-1}(\boldsymbol{s}_{t-1},\boldsymbol{a}_{t-1})$. To maintain the comfortable temperature range, the third part of $R_t$ can be computed by $-C_{3,t}(\boldsymbol{s}_{t})$, where
\begin{equation}\label{f_11}
C_{3,t}(\boldsymbol{s}_{t})=({\left[ {{T_t} - {T^{\max }}} \right]^ + } + {\left[ {{T^{\min }} - {T_t}} \right]^+}),~\forall~t,
\end{equation}
which means that $C_{3,t}=0$ if $T^{\min}\leq T_t\leq T^{\max}$. Otherwise, $C_{3,t}=T_t-T^{\max}$ if $T_t>T^{\max}$, and $C_{3,t}=T^{\min}-T_t$ if $T_t<T^{\min}$.

Taking three parts into consideration, the final reward function can be designed as follows,
\begin{equation}
R_t=-\beta(C_{1,t-1}(\boldsymbol{s}_{t-1},\boldsymbol{a}_{t-1})+C_{2,t-1}(\boldsymbol{s}_{t-1},\boldsymbol{a}_{t-1}))-C_{3,t}(\boldsymbol{s}_{t}), \nonumber
\end{equation}
where $\beta$ denotes a positive weight coefficient in $^oC/\$$.

\subsubsection{Action-Value Function}
When jointly controlling the ESS and the HVAC system at time slot $t$, the HEMS agent intends to maximize the expected return it receives over
the future. In particular, the return is defined as the sum of the discounted rewards\cite{Sutton2018}, i.e., $R=\sum\nolimits_{i = 1}^\infty \gamma^{i-1}R_{t+i}$. Let $Q_{\pi}(\boldsymbol{s},\boldsymbol{a})$ be the action-value function under a policy $\pi$ (note that a policy is a mapping from states to probabilities of selecting each possible action), which represents the expected return if action $\boldsymbol{a}_{t}=\boldsymbol{a}$ is taken in state $\boldsymbol{s}_{t}=\boldsymbol{s}$ under the policy $\pi$. Then, the optimal action-value function $Q^*(\boldsymbol{s},\boldsymbol{a})$ is $\max_{\pi}Q_{\pi}(\boldsymbol{s}_{t},\boldsymbol{a}_{t})$ and can be calculated by the following Bellman optimality equation in a recursive manner, i.e.,
\begin{equation}\label{f_12}
\begin{array}{l}
{Q^*}(\boldsymbol{s},\boldsymbol{a}) =\mathbb{E}[R_{t+1} + \gamma {\max _{{\boldsymbol{a}}'}}{Q^*}({{\boldsymbol{s}}_{t + 1}},{\boldsymbol{a}}')|{{\boldsymbol{s}}_t} = \boldsymbol{s},{{\boldsymbol{a}}_t} = \boldsymbol{a}].\\ \nonumber
\quad \quad \quad \quad  = \sum\nolimits_{\boldsymbol{s'},\boldsymbol{r}} P (\boldsymbol{s'},\boldsymbol{r}|\boldsymbol{s},\boldsymbol{a})[\boldsymbol{r} + \gamma {\max _{{\boldsymbol{a}}'}}{Q^*}(\boldsymbol{s'},{\boldsymbol{a'}}),
\end{array}
\end{equation}
where $\boldsymbol{s'}\in \mathcal{S}$, $\boldsymbol{r}\in \mathcal{R}$, $\boldsymbol{a'}\in \mathcal{A}$, and $P\in \mathcal{P}$.

To obtain $Q^*(\boldsymbol{s},\boldsymbol{a})$, system state transition probabilities $P (\boldsymbol{s'},\boldsymbol{r}|\boldsymbol{s},\boldsymbol{a})$ are required. Since indoor temperature in the smart home could be affected by many disturbances, it is difficult to accurately obtain state transition probabilities. To overcome this challenge, Q-learning methods could be used, which do not require the knowledge of state transition probabilities. To support the case with continuous system states, a function approximator could be adopted to estimate Q-function. When a neural network with weight $\theta$ is adopted as the non-linear function approximator, we refer it as Q-network. In \cite{Mnih2015}, a deep Q-network (DQN) algorithm was proposed, which can use experience replay and target network to ensure the stability of reinforcement learning methods when function approximators are adopted. However, DQN cannot be directly applied to the problem with continuous action spaces since it needs to discretize the action space and lead to an explosion of the number of actions. As a result, low computational efficiency, decreased performance, and the requirement of more training data would be incurred\cite{Gao2019}\cite{Lillicrap2016}.

\section{DDPG-based Energy Management Algorithm}\label{s4}
In this section, we first propose a DDPG-based energy management algorithm. Then, we analyze the computational complexity of the proposed algorithm.

\subsection{Algorithmic Design}

\begin{figure}[!htb]
\centering
\includegraphics[scale=0.783]{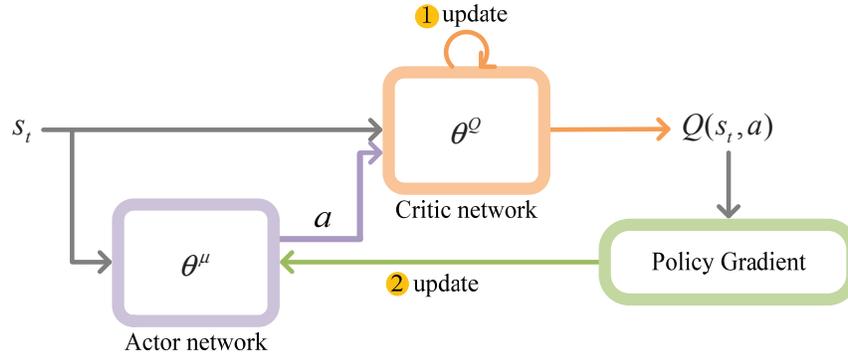}
\caption{Actor network and critic network in DDPG.}\label{fig_3}
\end{figure}

\begin{algorithm}[h]
\caption{The Proposed Energy management Strategy}
\label{alg_1}
\setcounter{AlgoLine}{0}
\LinesNumbered
\KwIn{System state $S_t$, testing time slots $H_{\text{test}}$}
\KwOut{System decision $\boldsymbol{a}_t=(f_t,e_t)$ in each time slot}

Load the weight of the actor network $\theta^{\mu}$ obtained by the Algorithm~\ref{alg_2}\;

\For{$t$=1,2,$\cdots$,$H_{\text{test}}$}
{
Select action $\boldsymbol{a}_{t}=\mu(\phi(\boldsymbol{s}_{t})|\theta^{\mu})$\;

Execute action $\boldsymbol{a}_{t}=(f_t,e_t)$ in smart home environment and observe next state $\boldsymbol{s}_{t+1}$ and reward $R_{t+1}$\;

}
\end{algorithm}

To solve the MDP problem defined in Section~\ref{s3_6}, we propose a DDPG-based energy management algorithm. Different from DQN, DDPG is capable of dealing with continuous states and actions. For example, just two network outputs are needed to represent continuous actions in this paper, which avoids the explosion of the number of actions. Since DDPG is a kind of actor-critic methods (i.e., methods that learn approximations to both policy function and value function), actor network and critic network are incorporated, which are shown in Fig.~\ref{fig_3}. The input and output of actor network is the environment state $\boldsymbol{s}_t$ and action $\boldsymbol{a}$, respectively. Then, $\boldsymbol{a}$ and $\boldsymbol{s}_t$ are adopted as the input of critic network, whose output is action-value function (i.e., $Q(\boldsymbol{s}_t,\boldsymbol{a})$). Next, the policy gradient can be computed and used to update the weight of actor network. Before computing $Q(\boldsymbol{s}_t,\boldsymbol{a})$, the weight of critic network should be updated based on two mechanisms, i.e., memory replay and target networks. More details will be introduced when explaining Algorithm 2.

\begin{algorithm}[h]
\caption{Training Deep Neural Networks with DDPG}
\label{alg_2}
\setcounter{AlgoLine}{0}
\LinesNumbered
\KwIn{Renewable generation output, non-shiftable power demand, outdoor temperature, electricity price}
\KwOut{The weights of actor network and critic network, i.e., $\theta^{\mu}$ and $\theta^{Q}$}
Initialize memory $\mathcal{D}$ of size $N$ \;

Initialize preprocess function $\phi(\boldsymbol{s}_{t})$\;

Randomly initialize critic network $Q(\phi(\boldsymbol{s}),\boldsymbol{a}|\theta^{Q})$ and actor network $\mu(\phi(\boldsymbol{s})|\theta^{\mu})$ with weights $\theta^{Q}$ and $\theta^{\mu}$, respectively\;

Initialize target networks $Q'$ and $\mu'$ by copying: $\theta^{Q'}\Leftarrow\theta^{Q}$, $\theta^{\mu'}\Leftarrow\theta^{\mu}$\;

\For{episode=1,2,$\cdots$,$M$}
{
   Receive the initial environment state $\boldsymbol{s}_0$\;

  \For{$t$=0,2,$\cdots$,$P-1$}
  {
    Select action $\boldsymbol{a}_{t}=\mu(\phi(\boldsymbol{s}_{t})|\theta^{\mu})+ \mathcal{N}_{t}$\;

    Execute action $\boldsymbol{a}_{t}$ in smart home environment and
    observe next state $\boldsymbol{s}_{t+1}$ and reward $R_{t+1}$\;

    Store $(\phi(\boldsymbol{s}_{t}),~\boldsymbol{a}_{t},~R_{t+1},~\phi(\boldsymbol{s}_{t+1}))$ in $\mathcal{D}$\;

    Sample a random mini-batch of $K$ transitions $(\phi(\boldsymbol{s}_{i}),~\boldsymbol{a}_{i},~R_{i+1},~\phi(\boldsymbol{s}_{i+1}))$ from $\mathcal{D}$,~$1\leq i\leq K$\;

    Set $y_{i} = R_{i+1}+ \gamma Q'(\phi(\boldsymbol{s}_{i+1}),\mu'(\phi(\boldsymbol{s}_{i+1})|\theta^{\mu'})|\theta^{Q'})$\;

    Update critic network by minimizing the loss:\;

    $L = {1\over K}\sum\nolimits_{i=1}^{K}{{{(y_{i}-Q(\phi(\boldsymbol{s}_{i}),~\boldsymbol{a}_{i}|\theta^{Q}))}^2}}$\;

    Update actor policy using sampled policy gradient:\;

    $\sum\nolimits_{i=1}^K{\frac{{\nabla_a}Q(\phi(\boldsymbol{s}),\boldsymbol{a}|\theta^{Q}){|_{\boldsymbol{s} ={\boldsymbol{s}_{i}},\boldsymbol{a}=\mu(\phi(\boldsymbol{s}_{i}))}}}{K}{\nabla_{\theta^{\mu}}}\mu (\phi(\boldsymbol{s})|\theta^{\mu}){|_{\boldsymbol{s}_{i}}}}$\;

    Update target networks:\;

    $\theta^{Q'} \leftarrow \tau \theta^{Q}+(1-\tau)\theta^{Q'}$\;

    $\theta^{\mu'} \leftarrow \tau \theta^{\mu}+(1-\tau)\theta^{\mu'}$\;

  }
}
\end{algorithm}

\begin{figure}
\centering
\subfigure[Actor network]{
\begin{minipage}[b]{0.6075\textwidth}
\includegraphics[width=1\textwidth]{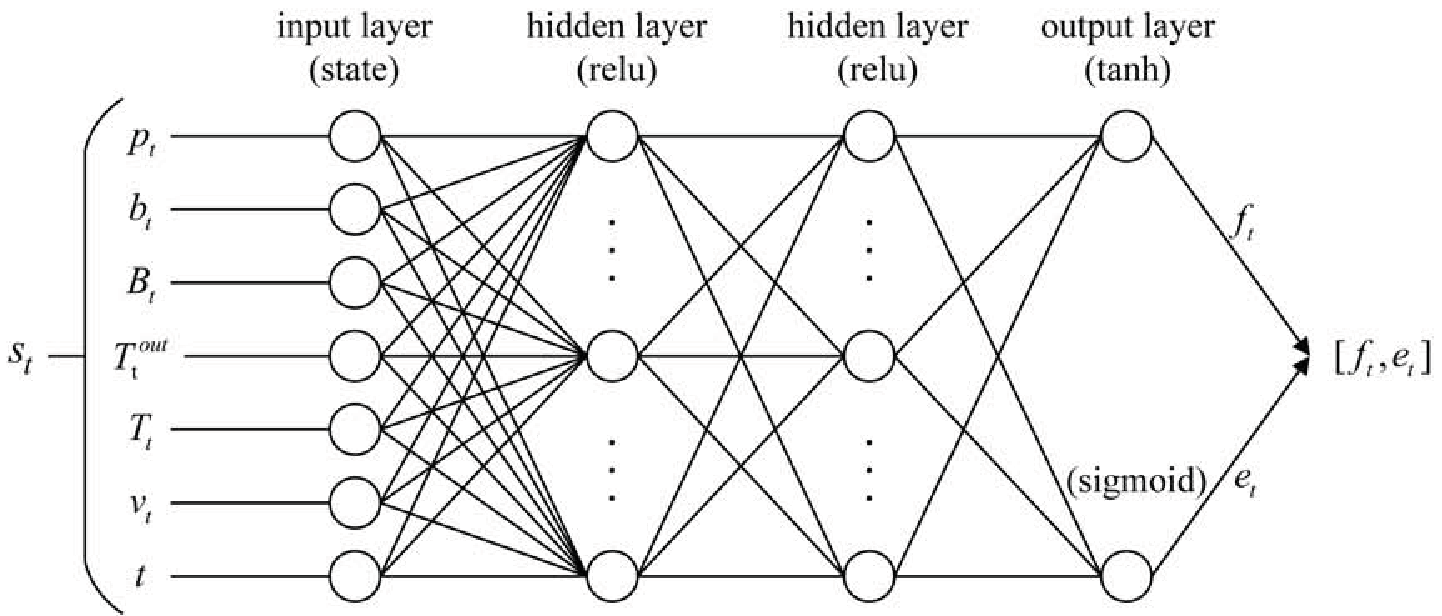}
\end{minipage}
}\\
\subfigure[Critic network]{
\begin{minipage}[b]{0.6075\textwidth}
\includegraphics[width=1\textwidth]{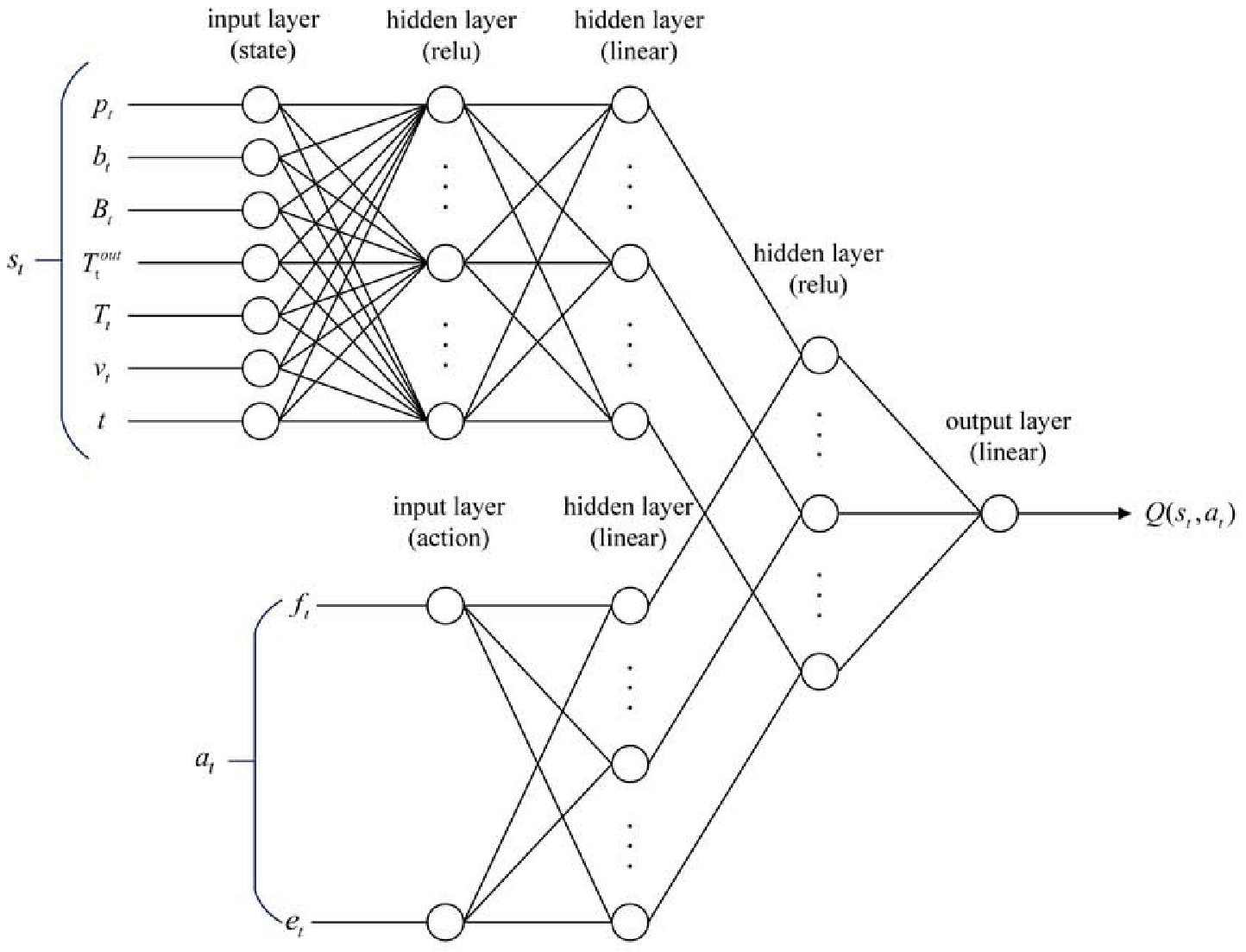}
\end{minipage}
}
\caption{The architectures of actor network and critic network.} \label{fig_4}
\end{figure}

The proposed DDPG-based energy management algorithm can be found in the Algorithm 1, where the key step is to load the weight of the actor network $\theta^{\mu}$, which is trained by Algorithm 2. In each time slot, the actor network selects an action on ESS charging/discharing power and HVAC input power according to the current environment state $\boldsymbol{s}_t$. Then, the action $\boldsymbol{a}_t$ is executed and the environment state becomes $\boldsymbol{s}_{t+1}$. Meanwhile, the reward $R_{t+1}$ is obtained. In Algorithm 2, we first initialize a replay memory $\mathcal{D}$ with capacity $N$, which stores the transition tuple $(\boldsymbol{s}_{t},~\boldsymbol{a}_{t},~R_{t+1},~\boldsymbol{s}_{t+1})$. Moreover, a preprocess function $\phi(\boldsymbol{s}_{t})$ is introduced to facilitate the learning process by normalizing the input data. Specifically, each component in the environment state at time slot $t$ (e.g., $\kappa_t$) should be normalized within the range [0,1] using the following expression: $\frac{\kappa_t-\min_t{\kappa_t}}{\max_t{\kappa_t}-\min_t{\kappa_t}}$. Then, we randomly initialize critic network $Q(\phi(\boldsymbol{s}),\boldsymbol{a}|\theta^{Q})$ and actor network $\mu(\phi(\boldsymbol{s})|\theta^{\mu})$ with weights $\theta^{Q}$ and $\theta^{\mu}$, respectively. Their architectures in the proposed energy management algorithm are described by Fig.~\ref{fig_4}, where there are two hidden layers in the actor network and four hidden layers in the critic network. Next, we initialize the weights of target critic network $Q(\phi(\boldsymbol{s}),\boldsymbol{a}|\theta^{Q'})$ and target actor network $\mu(\phi(\boldsymbol{s})|\theta^{\mu'})$ by copying, i.e., $\theta^{Q'}\leftarrow\theta^{Q}$ and $\theta^{\mu'}\leftarrow\theta^{\mu}$. In each time slot of each episode, an action is selected based on the following expression in the line 8, i.e.,
\begin{equation}\label{f_15}
\boldsymbol{a}_{t}=\mu(\phi(\boldsymbol{s}_{t})|\theta^{\mu})+ \mathcal{N}_{t},
\end{equation}
where $\mathcal{N}_{t}$ is the exploration noise. In this paper, we use the following way to introduce exploration noise, i.e.,
\begin{equation}\label{f_16}
\boldsymbol{a}_{t} = \left\{ \begin{array}{l}
\mu (\phi ({{\bf{s}}_t})|{\theta ^\mu }),~~\;if\;\omega_t  > \xi_t, \\
({U_{t,1}},{U_{t,2}}),\quad \,if\;\omega_t  \le \xi_t, \;
\end{array} \right.
\end{equation}
where $\omega_t$, $U_{t,1}$, and $U_{t,2}$ follow uniform distributions with parameters (0,1), (-$d^{\max}/\max\{c^{\max},d^{\max}\}$,~$c^{\max}/\max\{c^{\max},d^{\max}\}$), and (0,1), respectively. $\xi_t=\max(\xi_t-\zeta*(episode-N/P),\xi_{\min})$, $\xi_0=1$ and $0<\zeta<1$. After $\boldsymbol{a}_{t}$ is obtained, it will be applied to ESS and the HVAC system. At the end of time slot $t$, the new state $\boldsymbol{s}_{t+1}$ and the reward $R_{t+1}$ are returned from the environment. Then, the transition tuple $(\phi(\boldsymbol{s}_{t}),~\boldsymbol{a}_{t},~R_{t+1},~\phi(\boldsymbol{s}_{t+1}))$ will be stored in the memory for the training of actor and critic networks as shown in the line 10. Next, $K$ transitions are randomly sampled for training deep neural networks, i.e., actor network, critic network, target actor network, and target critic network. As shown in lines 12-14, $Q(\phi(\boldsymbol{s}_{i}),~\boldsymbol{a}_{i})$ and $y_i$ generated by critic network and target network are used to calculate mean square error loss. By minimizing the loss function, the weight of critic network could be updated. Then, we can calculate the sampled policy gradient as shown in the line 15, which is used to update the weight of actor network. Finally, the weights of target actor network and target critic network could be updated as shown in lines 17-19. Note that a small $\tau$ should be selected in order to improve the learning stability. Typically, $0<\tau\ll 1$.

\subsection{Algorithmic Computational Complexity}
In Algorithm 1, it can be observed that the computational complexity of the proposed energy management algorithm depends on the number of testing slots $H_{\text{test}}$. Since simple calculations are carried out in Algorithm 1, its computational complexity can be described by $\mathcal{O}(H_{\text{test}})$. Given the fixed testing time horizon, a shorter duration of a time slot would results in a larger $H_{\text{test}}$. However, the time slot's duration can not be selected arbitrarily in practice due to the following reasons. On one hand, too long duration would results in the loss of many control opportunities of saving energy cost and maintaining a comfortable temperature range. On the other hand, too short duration may affects the training convergence of DRL-based algorithms since the control actions taken by the DRL agent cannot take effect immediately in terms of environment states (e.g., indoor temperature)\cite{Zhang2018}. Therefore, the duration of a time slot should be selected appropriately in practice. In existing works, the typical duration of a time slot is several minutes or one hour (e.g., 15 minutes\cite{Wei2017}, 1 hour\cite{Zhang2018}), which is far greater than the computation time of the proposed energy management algorithm in a time slot. Therefore, the proposed energy management algorithm can be implemented in a real-time way.

\section{Performance Evaluation}\label{s5}
In this section, we evaluate the performance of the proposed energy management algorithm. We first describe the simulation setup. Then, we describe the baselines used for performance comparisons. Finally, we provide simulation results about algorithmic convergence process, algorithmic performance under varying $\beta$, algorithmic effectiveness, and algorithmic scalability.

\subsection{Simulation setup}
In simulations, we use real-world traces related to solar generation, non-shiftable power demand, outdoor temperature, and electricity price, which are extracted from Pecan Street database\footnote{https://www.pecanstreet.org/}. Note that such database is the largest real-world open energy database on the planet and includes the data related to home energy consumption and solar generation of the Mueller neighborhood in Austin, Texas, USA. For simplicity, the cooling mode of a residential HVAC system is considered. Since summers in Austin are very hot\footnote{https://en.wikipedia.org/wiki/Austin,\_Texas\#Climate}, we use the data during the period from June 1 to August 31, 2018 for model training and testing. To be specific, the data in June and July is used to train neural network models and the data in August is adopted for performance testing. Some important system parameters are configured as follows: $u_t=0.9v_t$\cite{liangyu2015}, $\gamma=0.995$, $\eta_{\text{c}}=\eta_{\text{d}}=0.95$\cite{Xu2010}, $\zeta=0.0005$, $\xi_{\min}=0.1$, $T^{\min}=66.2^oF (19^oC)$\cite{Wei2017}, $T^{\max}=75.2^oF (24^oC)$\cite{Wei2017}, other parameter configurations are shown in TABLE~\ref{table_1}, where $\alpha_a$ and $\alpha_c$ denote the learning rate of actor network and critic network, respectively. In TABLE~\ref{table_1}, $N_a$ and $N_c$ denote the number of neurons in each hidden layer of actor network and critic network, respectively. To simulate the environment, we adopt the following indoor temperature dynamics model for simplicity, i.e., $T_{t+1}=\varepsilon T_{t}+(1-\varepsilon)(T_t^{\text{out}}-\frac{\eta_{\text{hvac}}}{A}e_t)$\cite{Zhang2016}\cite{Pilloni2018}\cite{Constantopoulos1991}\cite{Thatte2012}, where $\varepsilon=0.7$\cite{Deng2016}, $\eta_{\text{hvac}}=2.5$\cite{Constantopoulos1991}, $A=0.14 kW/^{o}F$\cite{Constantopoulos1991}. Note that the variant of the proposed energy management algorithm can be applicable to any indoor temperature dynamics model by incorporating more environment-related variables in system state, e.g., relative humidity and solar radiation intensity.

\begin{table}[!htb]
\renewcommand{\arraystretch}{1.3}
\caption{Main parameter settings} \label{table_1} \centering
\begin{tabular}{|c|c|c|c|}
\hline  $H_{\text{test}}$      &744~hours &$\Delta t$           &1~hour\\
\hline  $B^{\max}$             &6$kWh$    &$B^{\min}$           &0.6$kWh$\\
\hline  $B_{\text{0}}$         &1.2$kWh$  &$c^{\max}$           &3$kW$\\
\hline  $d^{\max}$             &3$kW$     &$e^{\max}$           &2$kW$\\
\hline  $M$                    &3000      &$P$                  &24  \\
\hline  $K$                    &120       &$N$                  &$24000$\\
\hline  $\alpha_a$             &0.0001    &$\alpha_c$           &0.001\\
\hline  $N_a$                  &300,600   &$N_c$                &300,600,600,600 \\
\hline  $\tau$                 &0.001     &Optimizer            &Adam \\
\hline
\end{tabular}
\end{table}

\subsection{Baselines}
To evaluate the performance of the proposed algorithm, we adopt three baselines as follows.
\begin{itemize}
  \item \emph{Baseline1}: this scheme adopts ON/OFF policy\cite{Wei2017} for building HVAC control but without considering the use of the ESS. Specifically, the HVAC system will be turned on if $T_i>T^{\max}$ and it will be turned off if $T_i<T^{\min}$.
  \item \emph{Baseline2}: this scheme uses the DDPG-based control policy in this paper for HVAC control but without considering the use of the ESS, i.e., $c^{\max}=d^{\max}=0$. Based on the performance comparison between \emph{Baseline2} and the proposed algorithm, the energy cost saving caused by the use of the ESS can be known. Similarly, the energy cost saving incurred by the use of DDPG-based control policy can be obtained by comparing the performance of \emph{Baseline2} with that of \emph{Baseline1}.
  \item \emph{Baseline3}: this scheme intends to minimize the cumulative cost during the testing period $H_{\text{test}}$ (i.e., $\sum\nolimits_{t=1}^{H_{\text{test}}}(C_{1,t}+C_{2,t})$) with the consideration of constraints (1)-(8), assuming that all uncertainty system parameters and the dynamics model of indoor temperature can be known beforehand. Although the optimal solution of this scheme is not achievable in practice due to the existence of parameter and model uncertainties, it can provide the lower bound for the performance of the proposed algorithm when all constraints in \textbf{P1} are satisfied.
\end{itemize}

\subsection{Simulation Results}

\subsubsection{Algorithmic convergence process}

\begin{figure}[!htb]
\centering
\includegraphics[scale=0.675]{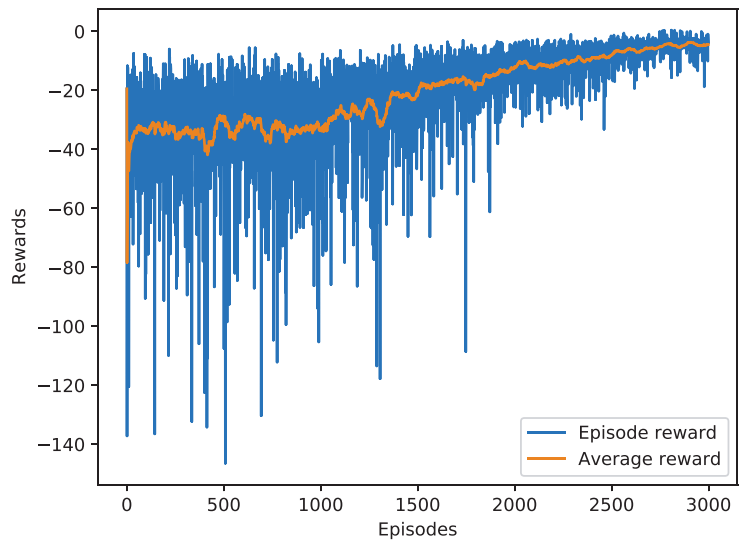}
\caption{The convergence process of the Algorithm 2.}\label{fig_5}
\end{figure}

According to Algorithm 1, the proposed energy management algorithm needs to know the training result of Algorithm 2 before testing. In Fig.~\ref{fig_5}, the reward received during each episode generally increases. Since the minimum exploration probability $\xi_{\min}$ is 0.1 and system parameters (e.g., solar radiation power, non-shiftable power demand, outdoor temperature, and electricity price) are varying in each episode, the episode reward fluctuates within a small range. To show the changing trend of rewards more clearly, we provide the average value of the past 50 episodes. In Fig.~\ref{fig_5}, it can be found that the average reward generally increases and becomes more and more stable.

\subsubsection{Algorithmic performance under varying $\beta$}

\begin{figure}
\centering
\subfigure[Total energy cost]{
\begin{minipage}[b]{0.54\textwidth}
\includegraphics[width=1\textwidth]{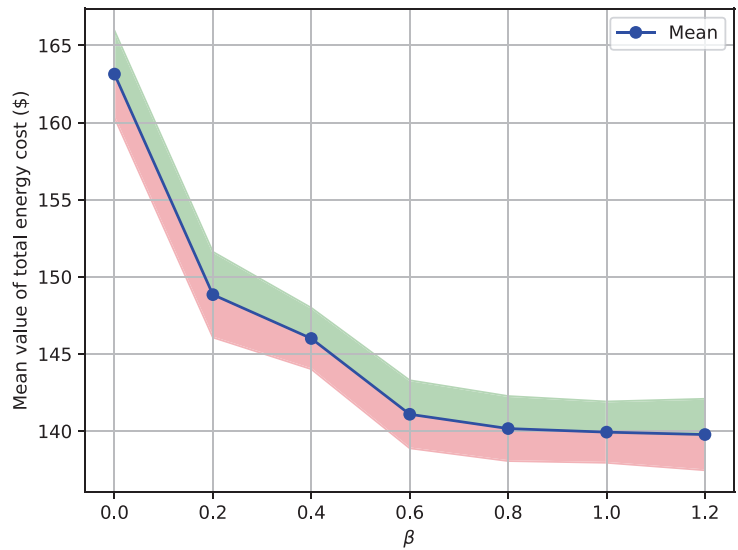}
\end{minipage}
}\\
\subfigure[Total temperature deviation]{
\begin{minipage}[b]{0.54\textwidth}
\includegraphics[width=1\textwidth]{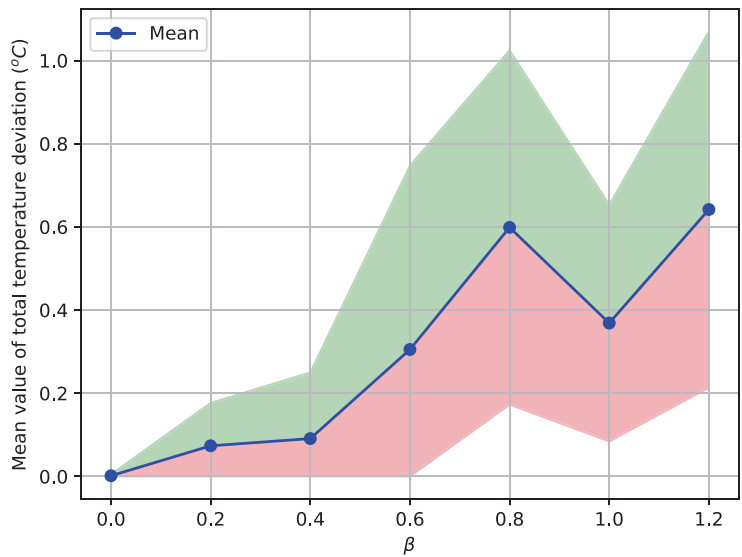}
\end{minipage}
}
\caption{The impact of $\beta$ on the performance of the proposed algorithm.} \label{fig_6}
\end{figure}

Since many random number generators are adopted in neural network initialization, mini-batch data collection for training, and action choice, the performance of the proposed algorithm is varying even the same system parameters are configured. To show the impact of $\beta$ on the performance of the proposed algorithm more clearly, mean values of total energy cost (i.e., the sum of energy cost and ESS depreciation cost) and total temperature deviation with 95\% confidence interval across 40 runs are considered and the corresponding results can be found in Fig.~\ref{fig_6}. It can be observed that the mean value of total energy cost and that of total temperature deviation generally decreases and increases with the increase of $\beta$, respectively. Such tendency is obvious since larger $\beta$ results in more importance of energy cost and less importance of temperature deviation. By taking mean values of total energy cost and total temperature deviation into consideration, a proper value of $\beta$ is 1 when the mean value of total temperature deviation is less than $1^oC$.

\begin{figure}
\centering
\subfigure[Mean value of total energy cost]{
\begin{minipage}[b]{0.486\textwidth}
\includegraphics[width=1\textwidth]{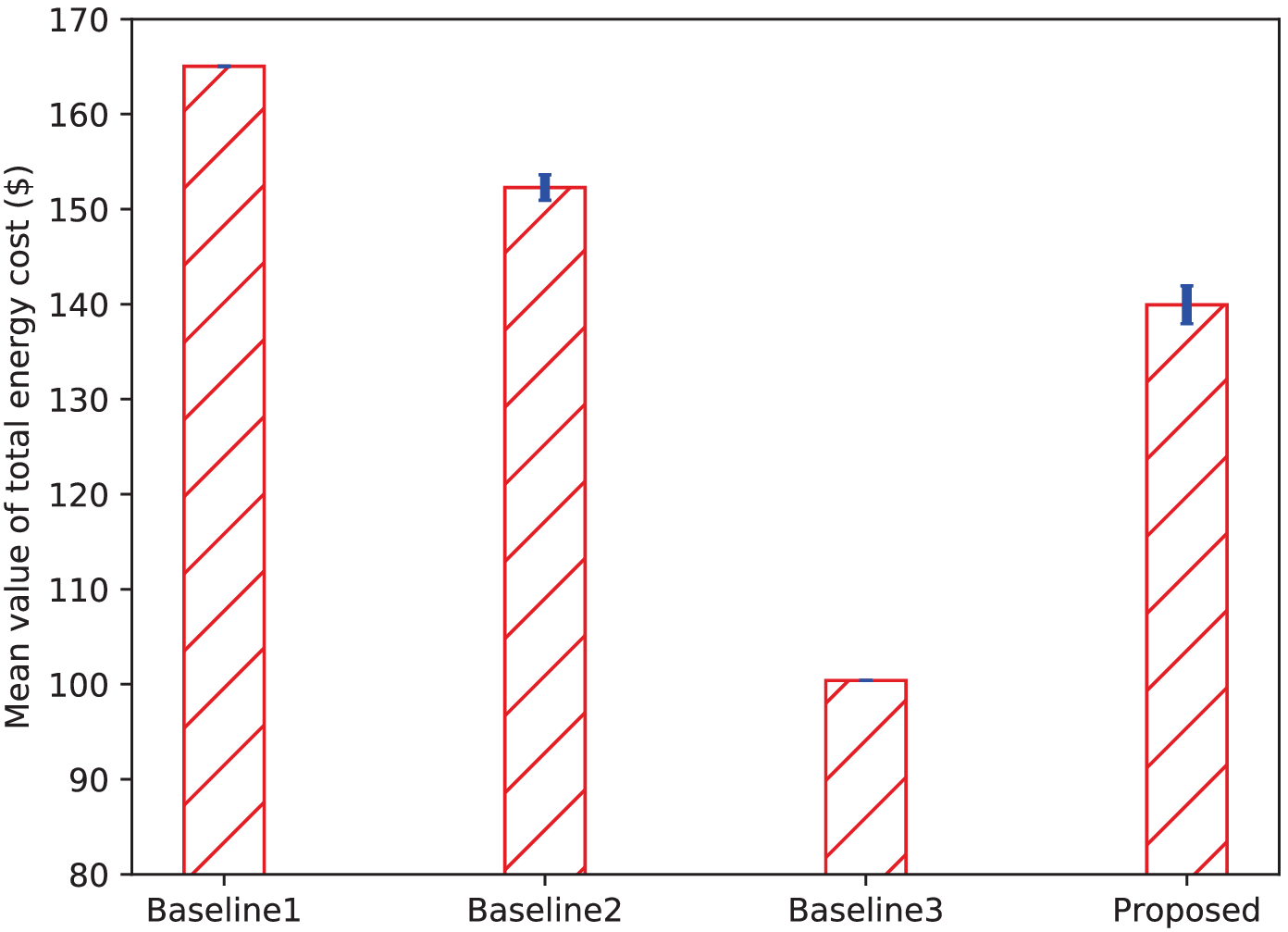}
\end{minipage}
}\\
\subfigure[Mean value of total temperature deviation]{
\begin{minipage}[b]{0.486\textwidth}
\includegraphics[width=1\textwidth]{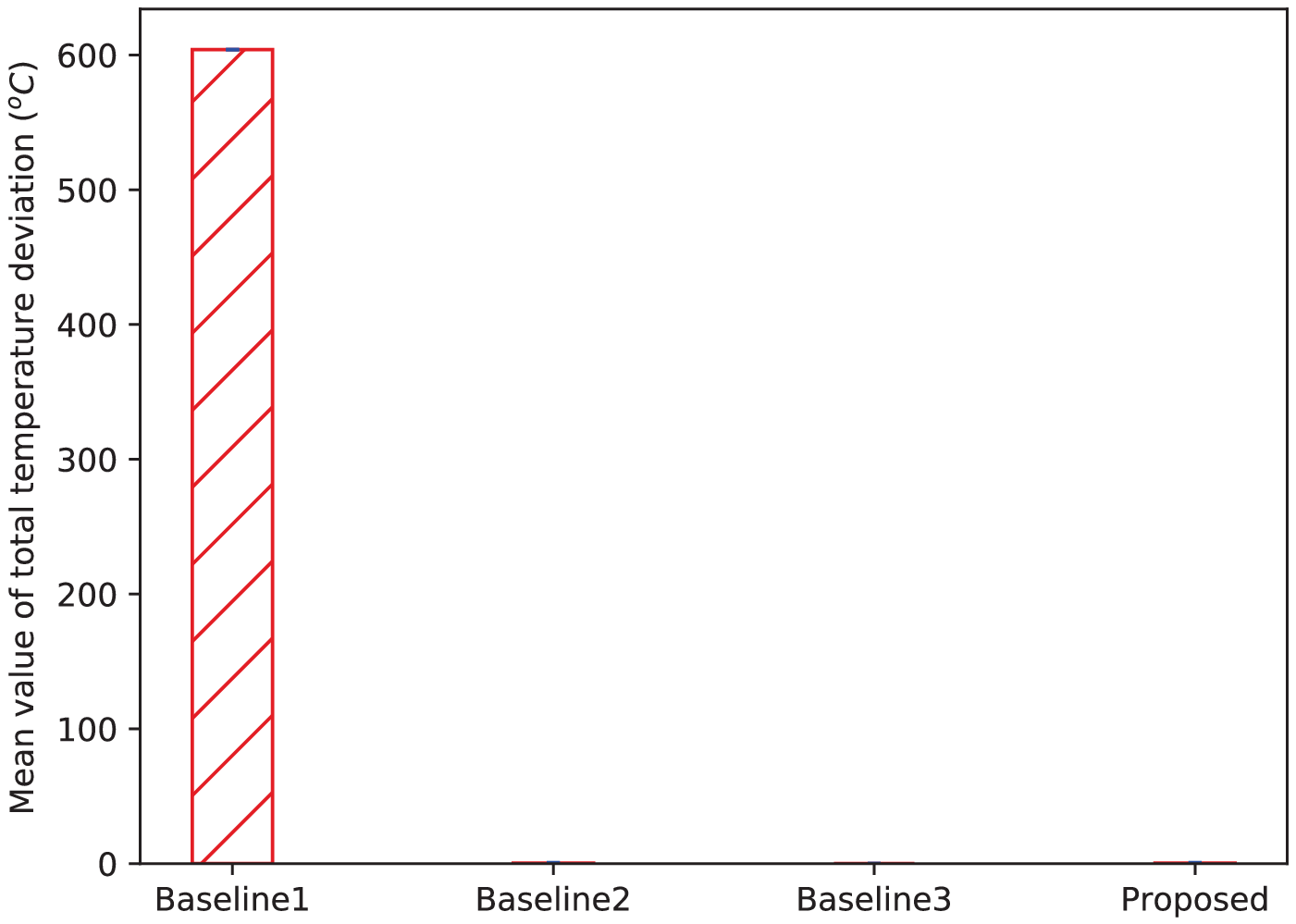}
\end{minipage}
}\\
\subfigure[Indoor temperature]{
\begin{minipage}[b]{0.486\textwidth}
\includegraphics[width=1\textwidth]{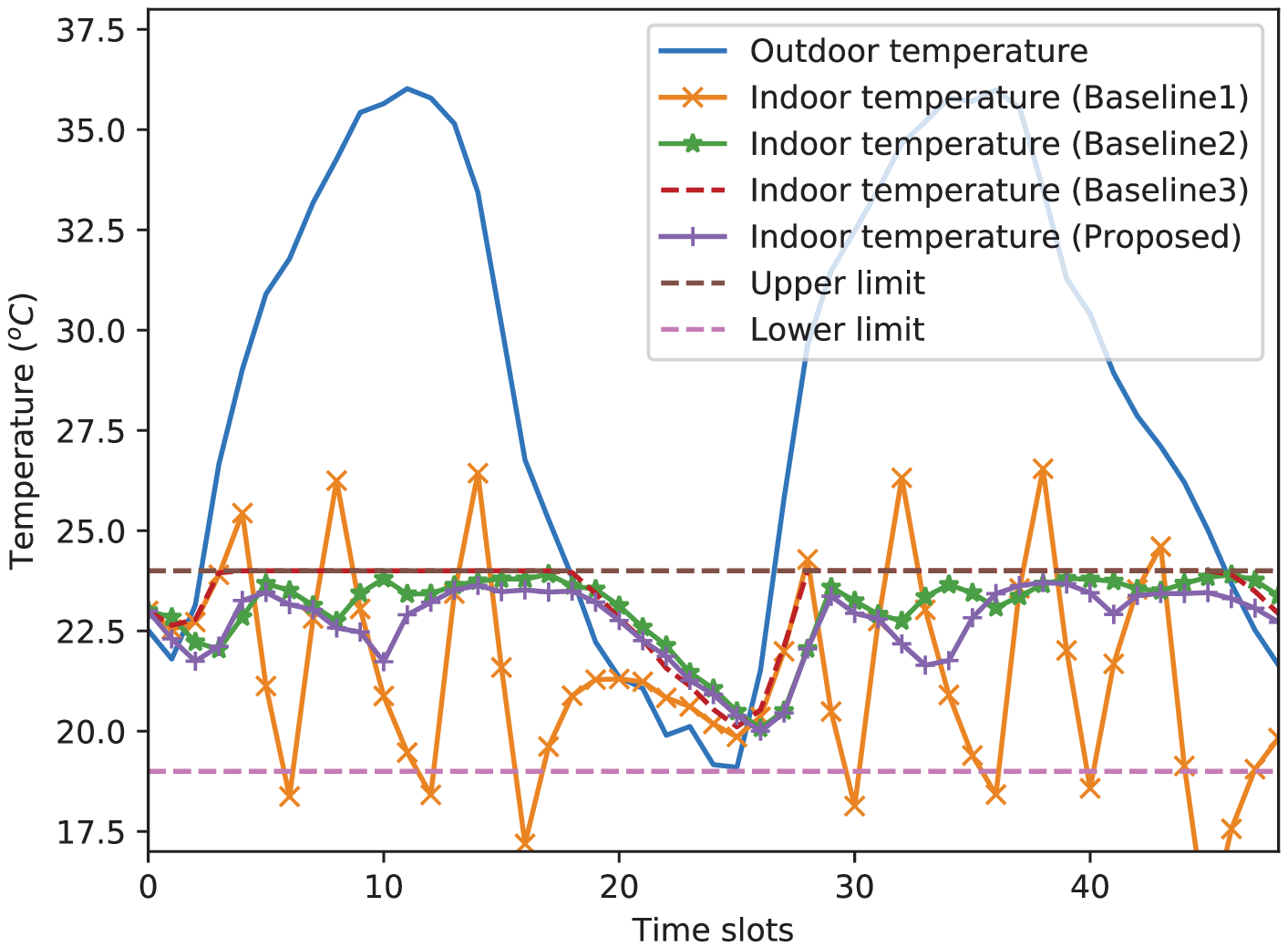}
\end{minipage}
}
\caption{Performance comparisons among three schemes ($\beta=0.6$,~95\% confidence interval across 40 runs is considered).} \label{fig_7}
\end{figure}

\subsubsection{Algorithmic effectiveness}
\begin{figure}
\centering
\subfigure[Price]{
\begin{minipage}[b]{0.54\textwidth}
\includegraphics[width=1\textwidth]{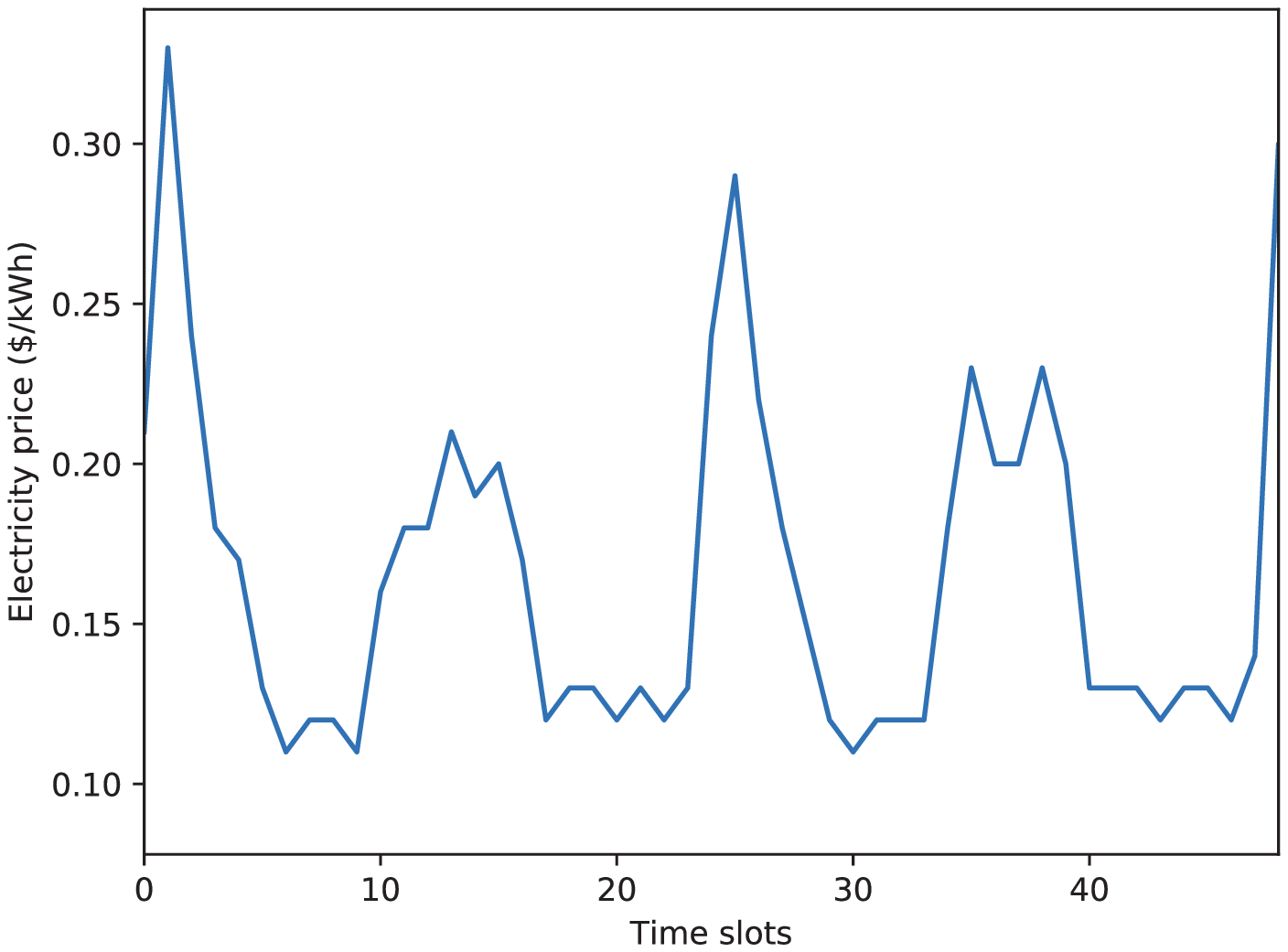}
\end{minipage}
}\\
\subfigure[HVAC input power]{
\begin{minipage}[b]{0.54\textwidth}
\includegraphics[width=1\textwidth]{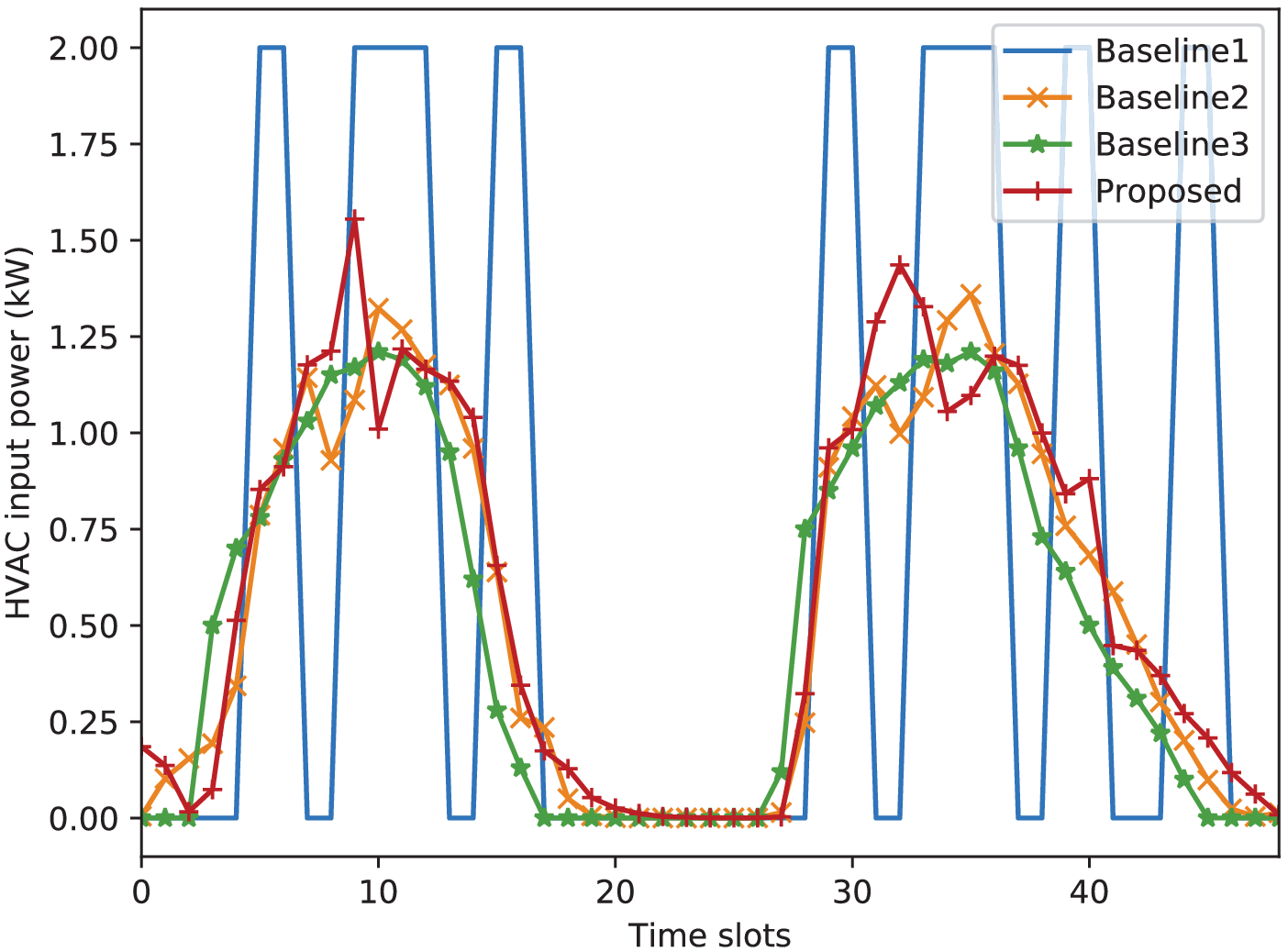}
\end{minipage}
}\\
\subfigure[ESS energy level]{
\begin{minipage}[b]{0.54\textwidth}
\includegraphics[width=1\textwidth]{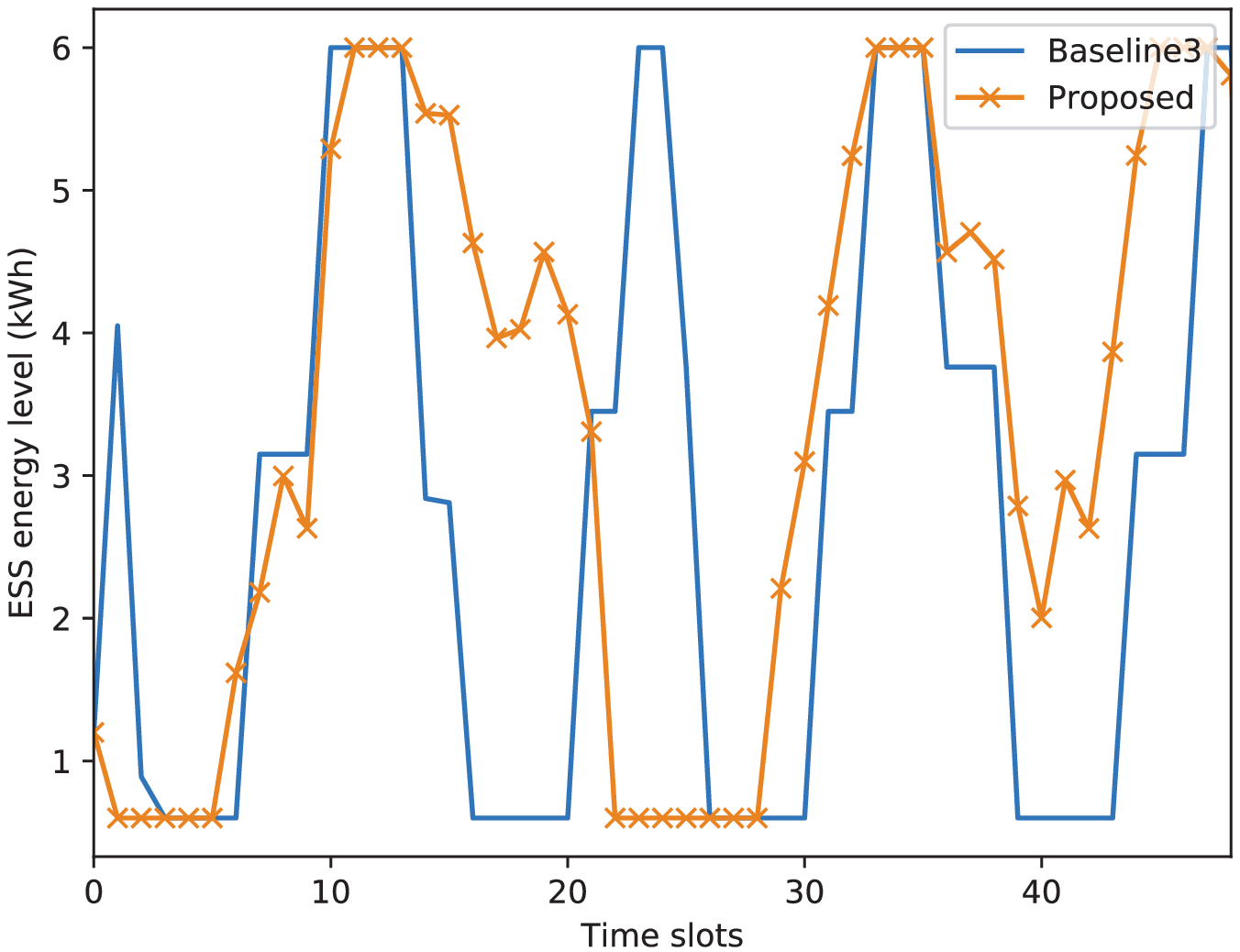}
\end{minipage}
}
\caption{Simulation results associated with ESS and HVAC systems.}\label{fig_8}
\end{figure}

Performance comparisons among four schemes are shown in Fig.~\ref{fig_7}, where the proposed energy management algorithm achieves better performance than \emph{Baseline1} and \emph{Baseline2}. To be specific, the proposed energy management algorithm can reduce the mean value of total energy cost by 15.21\% and 8.10\% when compared with \emph{Baseline1} and \emph{Baseline2}, respectively. Moreover, the mean value of total temperature deviation under the proposed algorithm is smaller than \emph{Baseline1} and \emph{Baseline2}, which can be illustrated by Figs.~\ref{fig_7}(b) and (c). Compared with \emph{Baseline1}, \emph{Baseline2} and the proposed algorithm could save energy cost by increasing/decreasing HVAC input power when electricity price is low/high, which can be depicted by Figs.~\ref{fig_8}(a) and (b). Compared with \emph{Baseline2}, the proposed algorithm could reduce energy cost by charging/discharging ESS when electricity price is low/high, which can be shown in Figs.~\ref{fig_8}(a) and (c). Though \emph{Baseline3} achieves the best performance, it requires all prior knowledge of uncertain system parameters and thermal dynamics model. Thus, \emph{Baseline3} is just adopted for performance reference. By observing the performance gap between the proposed algorithm and \emph{Baseline3}, it can be known that the potential of reducing the mean value of total energy cost is great. In future work, more training data and advanced DRL-based energy management algorithms would be adopted for reducing the performance gap.

\subsubsection{Algorithmic robustness}
\begin{figure}
\centering
\subfigure[Mean value of total energy cost]{
\begin{minipage}[b]{0.54\textwidth}
\includegraphics[width=1\textwidth]{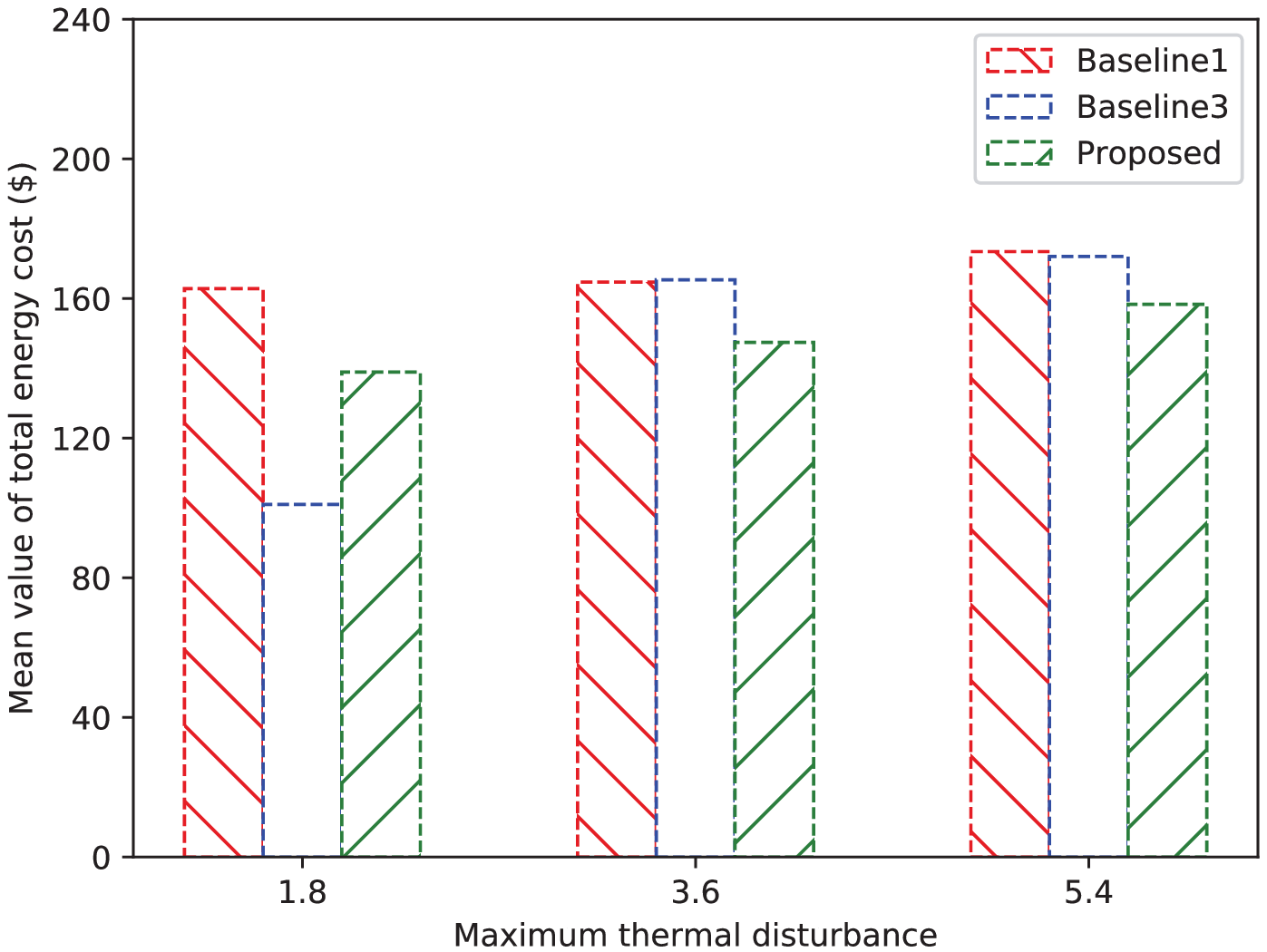}
\end{minipage}
}\\
\subfigure[Mean value of total temperature deviation]{
\begin{minipage}[b]{0.54\textwidth}
\includegraphics[width=1\textwidth]{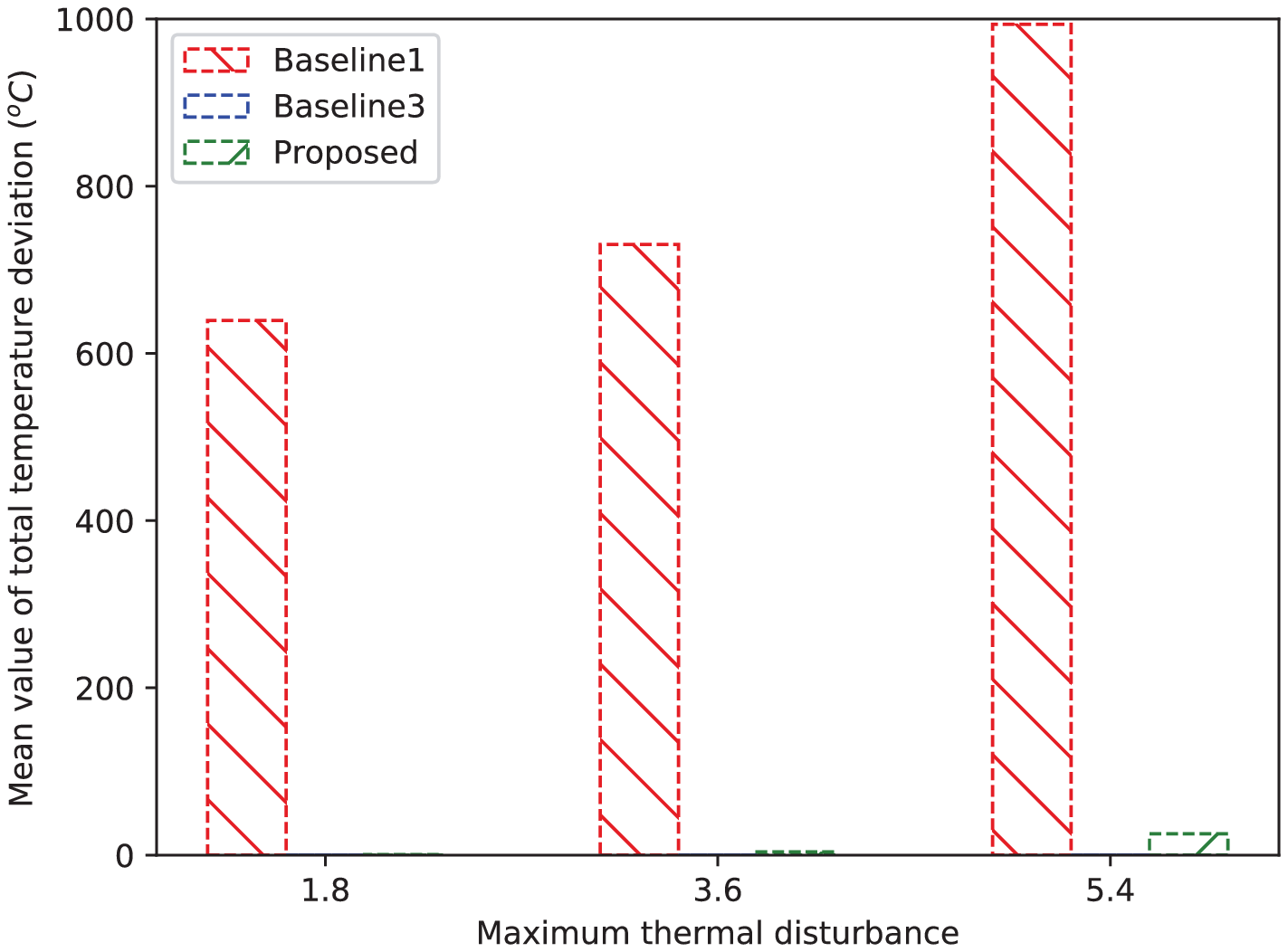}
\end{minipage}
}
\caption{The robustness of the proposed algorithm.} \label{fig_9}
\end{figure}

Note that the thermal dynamics model used in above-mentioned simulations can not capture thermal disturbances in practice, e.g., thermal disturbances from solar irradiance, lighting systems, and computers. Thus, we evaluate the robustness of the proposed algorithm when random thermal disturbance is introduced. To be specific, $T_{t+1}=\varepsilon T_{t}+(1-\varepsilon)(T_t^{\text{out}}-\frac{\eta_{\text{hvac}}}{A}e_t)+\epsilon_t$\cite{Kara2015}, where the error item $\epsilon_t$ is assumed to follow a uniform distribution with parameters $[\vartheta_l,~\vartheta_u]^oF$. In this scenario, three cases are considered, i.e., $\vartheta_u=-\vartheta_l={1.8,3.6,5.4}$. In Fig.~\ref{fig_9}, it can be observed that the proposed algorithm achieves better performances than \emph{Baseline1} under three cases. Compared with \emph{Baseline3}, the proposed algorithm can save the total energy cost by up to 10\% with a small increase of the total temperature violation. Moreover, unlike \emph{Baseline3}, the proposed algorithm does not require any prior knowledge of all uncertain parameters and thermal dynamics model. Therefore, the proposed algorithm has the potential of providing a more efficient and practical tradeoff between maintaining thermal comfort and reducing energy cost than \emph{Baseline3}.

\section{Conclusion}\label{s6}
In this paper, we proposed a DDPG-based energy management algorithm for a smart home to efficiently control HVAC systems and energy storage systems in the absence of a building thermal dynamics model, with the consideration of a comfortable temperature range and many parameter uncertainties. Extensive simulation results based on real-world traces showed the effectiveness and robustness of the proposed algorithm. In future work, more reasonable thermal comfort models and more types of controllable loads (e.g., electric vehicles, electric water heaters) would be incorporated. In addition, more opportunities of saving energy cost could be grasped by utilizing real-world occupant behavior information\cite{Guan2017}, which requires the adoption of more advanced deep neural network architectures/algorithms.

\end{document}